\documentclass[amsmath,amssymb,amsbsy,twocolumn,prb,showpacs,superscriptaddress]{revtex4}

\usepackage{amsfonts} 
\usepackage{amsmath}
\usepackage{amssymb}
\usepackage{cancel} 
\usepackage{bm}
\usepackage[usenames, dvipsnames]{color} 
\usepackage{dcolumn}
\usepackage{epic} 
\usepackage{epsfig}
\usepackage{feynmf}
\usepackage{graphicx}
\usepackage[breaklinks,colorlinks = true,linkcolor = red,urlcolor=cyan,citecolor=blue]{hyperref}
\usepackage{latexsym} 
\usepackage{mathrsfs}
\usepackage{soul}
\usepackage{subfigure}
\usepackage{times}
\usepackage{wrapfig} 
\usepackage{wasysym}
\usepackage{times}
\usepackage{xy} 

\newcommand{\La}{\line (1,0  ){12}}
\newcommand{\Lb}{\line (3,5 ){6}}

\newcommand{\Ld}{\line (-1,0){12}}
\newcommand{\Le}{\line (-3,-5){6}}

\newcommand{\C} {\circle*{4}}

\newcommand{\pA}{\put(-6,-10)}
\newcommand{\pB}{\put(6,-10)}
\newcommand{\pC}{\put(12,0)}

\newcommand{\pZ}{\put(0,0)}

\newcommand{\rhomb}{
  \pA{\C}\pB{\C}\pZ{\C}\pC{\C}
 }

\newcommand{\rhombH}{
  \begin{picture}(22,10)(-8,-6)
    \rhomb
    \pA{\La}\pC{\Ld}
  \end{picture}
}
\newcommand{\rhombV}{
  \begin{picture}(22,10)(-8,-6)
    \rhomb
    \pB{\Lb}\pZ{\Le}
  \end{picture}
}

\begin{document}

\title{$\mathbb{Z}_2$ topological liquid of hard-core bosons on a kagome lattice at $1/3$ filling}

\author{Krishanu Roychowdhury}
\affiliation{Max-Planck-Institut f\"ur Physik komplexer Systeme, Dresden 01187, Germany}
\author{Subhro Bhattacharjee}
\affiliation{Max-Planck-Institut f\"ur Physik komplexer Systeme, Dresden 01187, Germany}
\affiliation{International Centre for Theoretical Sciences, Tata Institute of Fundamental Research, Bangalore 560012, India}
\author{Frank Pollmann}
\affiliation{Max-Planck-Institut f\"ur Physik komplexer Systeme, Dresden 01187, Germany}

\begin{abstract}
We consider hard-core bosons on the kagome lattice in the presence of short range repulsive interactions and focus particularly on the filling factor $1/3$. In the strongly interacting limit, the low energy excitations can be described by the quantum fully packed loop coverings on the triangular lattice. Using a combination of tensor-product state based methods and exact diagonalization techniques, we show that the system has an extended $\mathbb{Z}_2$ topological liquid phase as well as a lattice nematic phase. The latter breaks lattice rotational symmetry. By tuning appropriate parameters in the model, we study the quantum phase transition between the topological and the symmetry broken phases. We construct the critical theory for this transition using a mapping to an Ising gauge theory that predicts the transition to belong to the $O(3)$ universality class. 
\end{abstract}


\maketitle
\section{Introduction}

Strong correlations in quantum many-body systems can yield novel phases of matter at low temperatures. The most prominent example is the fractional quantum Hall effect (FQHE)\cite{tsui1982two} which is characterized by its fractionalized excitations\cite{laughlin1983anomalous} and topological order\cite{laughlin1983anomalous,wen1990topological,wen1990ground-state} that manifests the long range quantum entanglement present in the underlying many-body ground state. While the FQHE requires strong magnetic fields, fractionalized states can occur more generally in correlated systems that preserve time-reversal symmetry. Following the seminal works by  Anderson\cite{anderson1973resonating, fazekas1974ground, anderson1987resonating} we  know that a class of two and three dimensional frustrated spin systems can realize  paramagnetic ground states, dubbed quantum spin-liquids (QSL), that support deconfined fractionalized $S=1/2$ (spinon) excitations as well as various forms of topological orders.\cite{wen1990topological, niu1990ground, wen1991mean, wen2004quantum}

Over the past decades, there has been a very active search for model systems stabilizing different types of topologically ordered states.\cite{wen1990topological, niu1990ground, wen1991mean, wen2004quantum} A fertile ground to realize such phases are frustrated systems in which the geometry of the lattice and/or competing interactions prohibit a simultaneous minimization of all the inter-particle interactions.\cite{wannier1950antiferromagnetism, diep2004frustrated} This can lead to the suppression of conventional (spontaneous symmetry breaking) orders and favor more exotic ground states. An associated question of much interest is about  quantum phase transitions between  topologically ordered  and  conventional symmetry breaking phases. In this regard, systems of hard-core bosons with mutual short range repulsions on various frustrated lattices have attracted much attention because they present a set of rich phase diagrams.\cite{heidarian2005persistent, isakov2006spin, isakov2006hard} A particularly interesting case is the kagome lattice where, for suitable fillings, a variety of numerical and analytical evidences now point to the existence of a topologically ordered $\mathbb{Z}_2$ liquid phase  over an extended parameter regime.\cite{balents2002fractionalization,sheng2005numerical} The low energy physics of such models in the strongly correlated limit  is generically  described by quantum dimer models (QDM)\cite{rokhsar1988superconductivity} on triangular lattices. This latter class of Hamiltonians is known to harbor points in their parameter space, the so called Rokhsar-Kivelson (RK) points,\cite{rokhsar1988superconductivity, moessner2001resonating, moessner2001phase} where  $\mathbb{Z}_2$ topological order is present and strong numerical evidence for its stability exists.\cite{moessner2001resonating} The intimate connection, on the other hand, between such hard-core boson models and $S=1/2$ models with XXZ anisotropy\cite{isakov2006hard} also makes these models relevant to research in quantum magnetism. With this insight, perhaps it is not surprising that there are proposals that the isotropic nearest neighbor Heisenberg antiferromagnet can potentially realize a $\mathbb{Z}_2$  topologically ordered ground state.\cite{yan2011spin, depenbrock2012nature}

In this paper, we explore the strong coupling physics of a hard-core boson model on kagome lattice with short range repulsive interactions particularly for the filling factor $f=1/3$. We show that the effective low energy theory is given by a QDM Hamiltonian on a triangular lattice with two dimers emanating from each site of the triangular lattice. This effective theory is thus equivalent to a quantum fully packed loop (FPL) model\cite{blote1994fully} on a triangular lattice. Such kind of models in the classical version have been extensively studied on some of the bipartite lattices.\cite{blote1989critical, blote1994fully, batchelor1994exact, kondev1995four, batchelor1996critical, jaubert2011analysis} The quantum FPL model on a square lattice has been studied by Shannon {\it et. al.} where they showed a correspondence to the XXZ model on the checkerboard lattice in certain {\it easy-axis} limits.\cite{shannon2004cyclic} 
Using a combination of numerical techniques (tensor product states formalism and exact diagonalization on clusters), we analyze this model on the triangular lattice. Our numerical analysis strongly advocates for a rich phase diagram of the quantum FPL on the triangular lattice consisting of an extended $\mathbb{Z}_2$ topological liquid phase as well as a crystalline phase, known as  lattice nematic (LN),\cite{mulder2010spiral} that breaks the three-fold rotational symmetry of the lattice (but not the translation symmetry). Taking clue from our numerical results, we then construct a critical theory for a continuous phase transition between the two phases. Unlike the usual theories of phase transition, this critical theory is not written in terms of the low energy long wavelength fluctuations of the LN order-parameter, but naturally in terms of ``fractionalized" Ising degrees of freedom sitting at the centers of the triangles of the kagome lattice. Mapping the problem to the language of Ising gauge theory,\cite{kogut1979introduction} we can isolate these critical degrees of freedom-- the so called {\it visons}\cite{senthil2001frac} (Ising magnetic flux\cite{read1989statistics}), whose condensation then describes the transition from the topological liquid to the LN. The order parameter is a bilinear in terms of the visons and hence, the above transition consists of an example of quantum criticality beyond the conventional Landau-Ginzburg-Wilson paradigm.\cite{senthil2004deconfined} Our calculation predicts that the critical theory belongs to the $O(3)$ universality class.

The remainder of the paper is organized as follows. We first introduce the model Hamiltonian and derive the effective model in Section \ref{sec_bos_mod}. We then present the numerical results and conclude about the phase diagram in Section \ref{sec_numerics}. In Section \ref{sec_trans} we derive an effective gauge theory description and discuss the nature of the transition between the liquid and the LN phase. We conclude with a brief summary in Section \ref{sec_summary}.


\section{Model Hamiltonian}
\label{sec_bos_mod}

We start by considering an extended Hubbard model of hard-core bosons on the sites of a kagome lattice  given by the Hamiltonian
\begin{align}
\mathcal{H}=\mathcal{H}_t+\mathcal{H}_V~,
\label{fullham}
\end{align}
where
\begin{eqnarray}
 \mathcal{H}_t =-t\sum_{\langle i,j \rangle} (b^{\dagger}_{i}b_{j} + \text{H.c.})
\label{hopping_gen}                 
\end{eqnarray}
describes the nearest-neighbor hopping, with amplitude $t$, for the hard-core bosons that are created (annihilated) by  $b^{\dagger}_i$ ($b^{\vphantom{\dag}}_{i}$) on the sites of the kagome lattice and
\begin{eqnarray}
 \mathcal{H}_V & = & V_1\sum_{\langle i,j \rangle} n_{i}n_{j} + V_2\sum_{\langle\langle i,j \rangle\rangle} n_{i}n_{j} +  V_3\sum_{\langle\langle\langle i,j \rangle\rangle\rangle} n_{i}n_{j}\nonumber\\
 &&~~~ - \tilde{\mu}\sum_{i} n_i
\label{potential_gen}                 
\end{eqnarray}
denotes respectively the first ($V_1$), second ($V_2$) and third ($V_3$) neighbor repulsive interactions among the bosons ($n_{i}=b_{i}^{\dag}b^{\vphantom{\dag}}_{i}$) along with a chemical potential $\tilde{\mu}$ that fixes the particle number. Interesting physics emerges at rational fractional fillings $p/q$ (with $p$ and $q$ being mutually prime and $p<q$).  We shall restrict ourselves to the specific fractional fillings of bosons which are $1/6$, $1/3$ and $1/2$.
 
%
%

\begin{figure}
\begin{centering}
\includegraphics[width=8.0truecm]{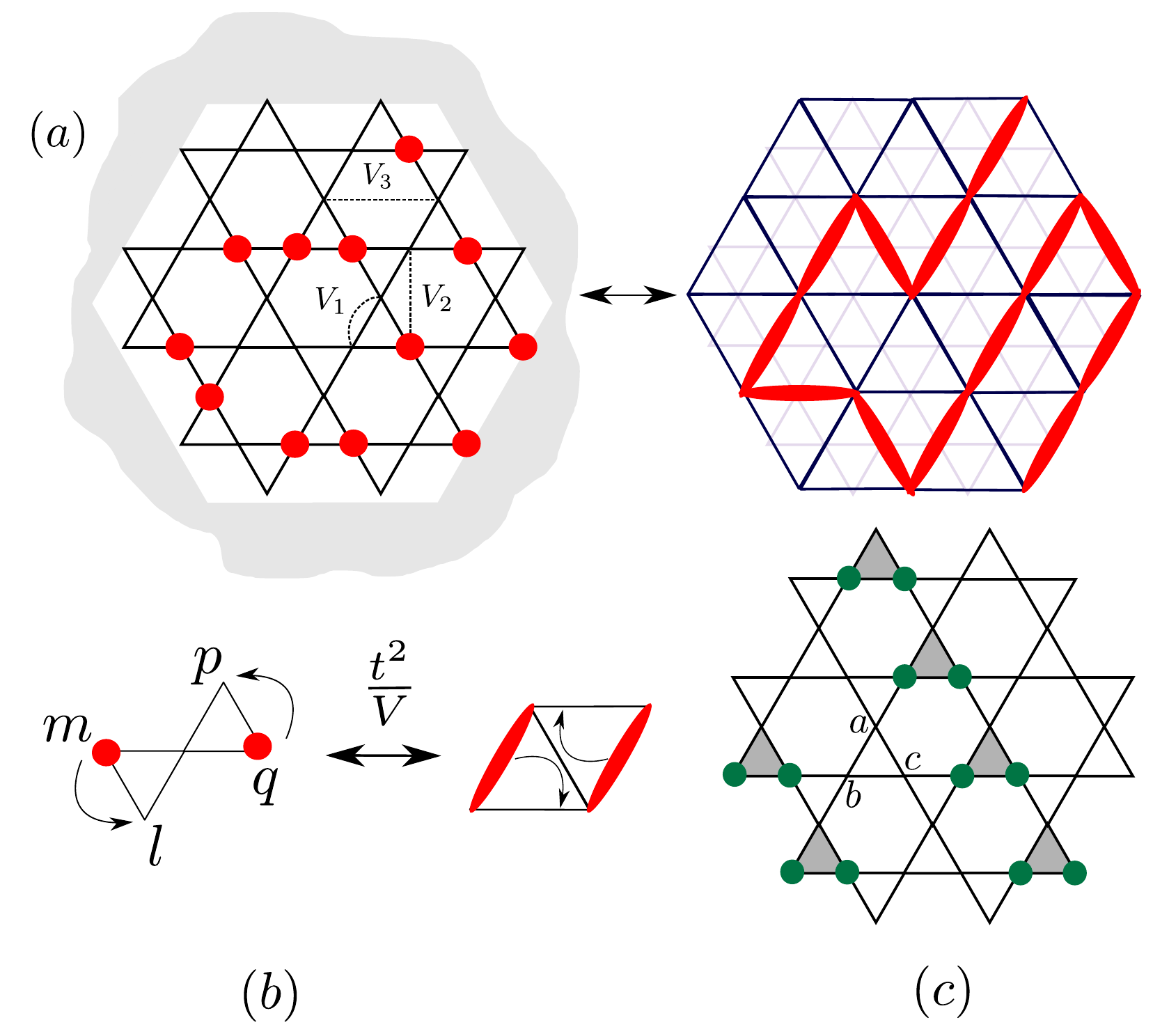}              
\end{centering}
\caption{(Color Online) (a) Mapping from the $1/3$ filled bosonic problem on kagome to FPL on triangular lattice. Each particle on the kagome lattice is mapped to a loop segment on the triangular lattice that is obtained by connecting the centers of the kagome hexagons. The strength of the 1st, 2nd and 3rd repulsive interactions are denoted as $V_1$, $V_2$ and $V_3$ respectively. The limit of $V_1=V_2=V_3=2V$ is the focus of this paper. Figure (b) demonstrates the allowed lowest order processes in $t/V$ (see the text). (c) Shown is the sublattice used for the gauge transformation that changes the sign of $g$ in Eq.~(\ref{h_eff}) as described in the text. The sublattice is constructed by the $b$ and $c$ sites of the shaded unit cells. 
\label{fig:loop}}
\end{figure}

At integer filling 0 or 1 there are only trivial product states possible (due to the hard-core constraint). At fractional filling factors and the presence of longer-range interactions much rich phase diagrams emerge. For example, at $1/3$ filling and strong nearest-neighbor repulsion ($V_1$), a plaquette ordered state\cite{isakov2006hard} is formed, while for $1/2$ filling a uniform superfluid persists for all values of $V/t$. When turning on further neighbor interactions given by $V_2$ and $V_3$, additional insulating lobes emerge at different fractional fillings.\cite{balents2005competing} Many of these bosonic insulators at fractional fillings can host interesting quantum phases with or without spontaneously broken symmetry. In the following, we  focus on the strong coupling phases occurring in the $1/2$, $1/3$ and $1/6$ lobes, with a particular focus on the $1/3$ lobe and show that topological as well as long ranged ordered phases can emerge. 

To this end we look at the strong coupling limit of the above Hamiltonian. For $t=0$, and $V_1=V_2=V_3=2V$, the interaction term can be written as
\begin{align}
 \mathcal{H}_{\text{V}} &
 = V\sum_{\left\{ \mbox{\small\hexagon} \right\}} \left[\left(n_{\mbox{\small\hexagon}}-\frac{\mu}{4V}\right)^2 - \frac{\mu^2}{16V^2}\right]~,
 \label{potential}                       
\end{align}
where $\mu= \tilde{\mu}+2V$ is the effective chemical potential and $n_{\mbox{\small\hexagon}}$ is the number of particles in each of the hexagons of the kagome lattice. 
It is clear from Eq.~(\ref{potential}) that for $\mu=4V, 8V, 12V$, $ \mathcal{H}_{\text{V}}$ is minimized by having $1, 2, 3$ bosons per hexagon respectively (or alternatively filling fraction of $f=1/6, 1/3$ and $1/2$ respectively). Clearly there are many different configurations of bosons that satisfy the above constraint, however, we note that since the hexagons share sites, the configurations for different hexagons are not completely independent. 

An insight to the number of states spanning the ground state sector of $\mathcal{H}_V$ for the above commensurate fillings can be obtained from the one-to-one correspondence between the ground state configurations of the bosons and the hard-core dimer coverings on the triangular lattice obtained by joining the centers of the hexagons of the kagome lattice as shown in Fig.~\ref{fig:loop}(a). 
Thus each site of the kagome lattice lies on a bond of the triangular lattice and the presence (absence) of a boson on that site can then be identified uniquely with the presence (absence) of a dimer on the corresponding bond of the triangular lattice. This immediately shows that at $1/6$ filling, the number of ground state bosonic configurations allowed by $\mathcal{H}_{V}$ is equivalent to the number of hard-core dimer coverings on the triangular lattice which is known to be extensive ($\sim 1.5^{N}$) in the system size, $N$ (an estimate\cite{PhysRevB.66.214513} based on Pauling's approximation\cite{pauling1935structure} gives $\sim 1.34^N$). Similarly $1/2$ filling can be cast into a 3-dimer (three non overlapping dimers emerging from each site) problem	m on the triangular lattice\cite{balents2002fractionalization} where Pauling estimate suggests that the extensive degeneracy of the ground state is $\sim 1.84^N$. 

In case of $1/3$ filling, which is the specific interest of the present paper, we obtain the equivalent fully-packed loop (FPL) model on the triangular lattice with two non-overlapping dimers emanating from each site and the dimers form non-intersecting loops as shown in Fig.~\ref{fig:loop}(a). Here the Pauling estimate shows that the number of loop configurations scales with system size as $\sim1.7^{N}$. Thus, in all the above cases, for $t=0$, as expected, the ground state is macroscopically degenerate and has a finite zero temperature entropy. Throughout the rest of this work, we shall  exploit the above equivalence between the bosons and the dimers and shall mostly use the language of the dimers, translating back to the bosons whenever applicable. 


\subsection{Effective model in the strong coupling limit}

Small but non-zero hopping ($t$) induces quantum fluctuations that (e.g. in the form of local ring exchange around small plaquettes\cite{thouless1965exchange, paramekanti2002ring, bernu2004multi, runge2004charge}) can lift the extensive ground state degeneracy by quenching the entropy of the classical model $(t=0)$  either by spontaneously breaking one or more symmetries of the system (quantum {\it order by disorder}\cite{villain1980order, shender1982orderbydisorder}) or more interestingly, by generating a long ranged quantum entangled state that does not break any symmetry of the Hamiltonian (quantum {\it disorder by disorder}\cite{PhysRevB.64.134424, PhysRevB.87.054404}) but can have non-trivial ``topological order". In this work, we shall show examples of both the routes taken by the bosons on the kagome lattice. 

To derive the effective Hamiltonian in the strong coupling limit, we treat $t/V$ as a small perturbation parameter to obtain (to the leading order in $t/V$)\cite{balents2002fractionalization} 
\begin{equation}
 \mathcal{H}_{\text{eff}} = -g\sum_{\alpha} (b^{\alpha\dagger}_{p}b^{\alpha\dagger}_{l}b_{m}^{\alpha}b_{q}^{\alpha} + \text{H.c})~,
\label{h_eff}
\end{equation}
where $g=t^2/V$ and $p$,$q$,$l$ and $m$ are the corner sites of the bow-tie labeled by $\alpha$ referring to Fig.~\ref{fig:loop}(b). 
The sum includes all bow-ties that are related by the $\mathcal{C}_3$ symmetry. 
The center of the bow tie may or may not be occupied  by a boson in case of $1/3$ and $1/2$ filling. 
Furthermore the sign of $g$ in Eq.~(\ref{h_eff}) can be altered by using a simple gauge transformation that multiplies all configurations with the factor $(-1)^{N_{\text{sub}}}$ where $N_{\text{sub}}$ is the number of particles in the sublattice shown in Fig.~\ref{fig:loop}(c).

%
%

\subsection{Dimer representation of the effective Hamiltonian}

\begin{figure}
\begin{centering}
\includegraphics[width=6.0truecm]{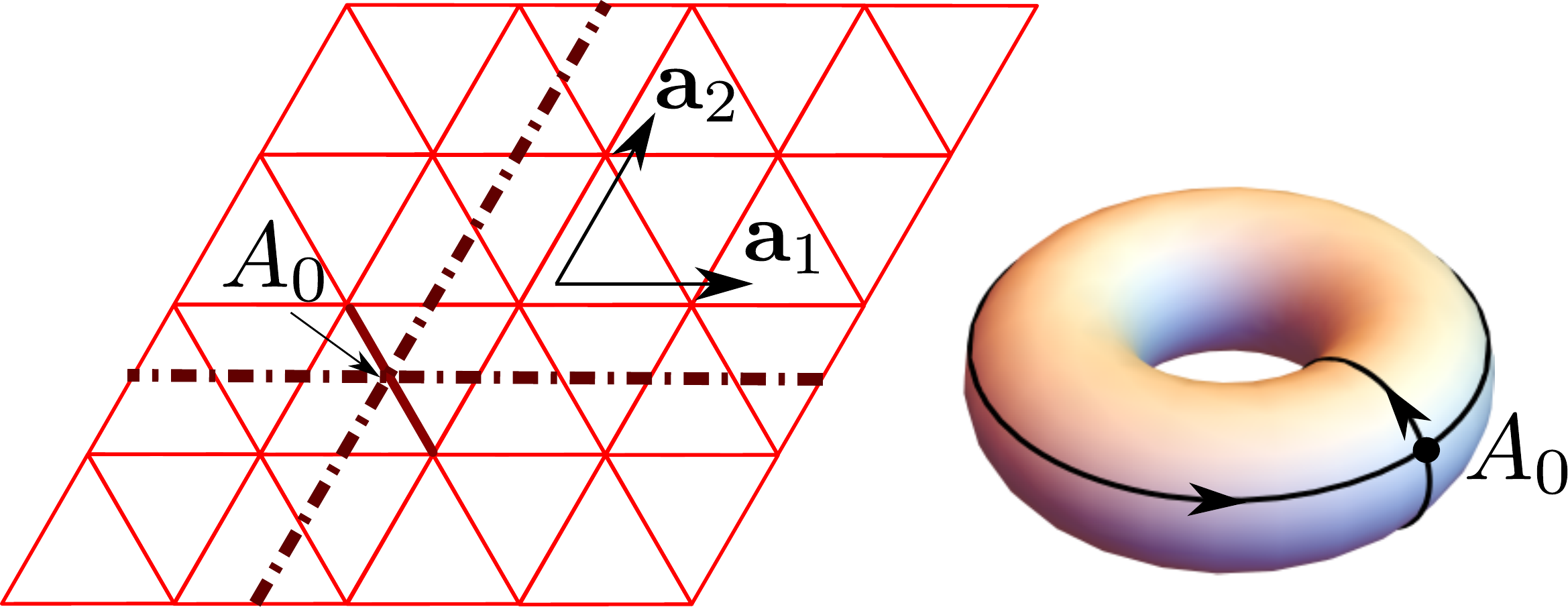}              
\end{centering}
\caption{(Color online) The triangular lattice is put on a torus by setting periodic boundary conditions along two independent directions given by the lattice vectors ${\bf{a}}_1$ and ${\bf{a}}_2$. $A_0$ denotes the intersection point (a link on the lattice) of the two non-contractible loops on the torus. There are four topological sectors characterized by the doublets ($0,0$), ($0,1$), ($1,0$) and ($1,1$).\cite{moessner2001resonating, ralko2005zero} The local Hamiltonian in Eq.~(\ref{tQDM}) can not mix configurations from different topological sectors hence, block diagonal in the full configuration basis.  
\label{fig:tor}}
\end{figure}

We now recast Eq.~(\ref{h_eff}) using the particle to dimer mapping mentioned in the previous section. 
In terms of the dimers, the effective Hamiltonian corresponds to the kinetic term of the QDM.\cite{moessner2001resonating}
\begin{equation}
  \mathcal{H}_{\text{eff}} = -g\sum_{\alpha} \left(\left|\rhombV\right>\left<\rhombH\right| +{\rm H.c.}\right)~,
\label{h_dimer}
\end{equation}
where the sum is over all rhombus shaped plaquettes ($\alpha$) on the triangular lattice. 

In addition to the kinetic term (above) the generic QDM also includes a potential term (known as Rokhsar-Kivelson (RK) potential),\cite{rokhsar1988superconductivity} which energetically favors  configurations with plaquettes having parallel dimers. Such potential terms are representative of higher order (four boson) terms that can be generated in the strong coupling expansion in $t/V$ of the underlying boson model. However, here we simply use this term as a free tuning parameter in the model.
The full Hamiltonian of the QDM is written as 
\begin{eqnarray}
  \mathcal{H}_{\text{RK}}&=&-g\sum_{\alpha} \left(\left|\rhombV\right>\left<\rhombH\right| +{\rm H.c.} \right) + \nonumber \\ 
  &V_{\text{RK}}& \sum_{\alpha}  \left(\left|\rhombV\right>\left<\rhombV\right| + \left|\rhombH\right>\left<\rhombH\right|\right),
\label{tQDM}
\end{eqnarray}
where positive (negative) $V_{\text{RK}}$ denotes repulsive (attractive) interaction between the parallel dimers in a given plaquette.
%
%

This generic form of the above Hamiltonian was first proposed on a square lattice by Rokhsar and Kivelson \cite{rokhsar1988superconductivity} (RK) in context of high temperature superconductivity. 
At a special point when $g=V_{\text{RK}}$, known as RK point, the spectrum contains zero energy ground state with the wave function given by the equal weight superposition of all allowed dimer configurations. 
The RK point features a $U(1)$ resonating valence bond (RVB) liquid phase with algebraic decay of dimer correlations.

However, on  non-bipartite lattices, the scenario changes drastically.\cite{moessner2001resonating, ralko2005zero} The RVB phase in this case fosters a gapped $\mathbb{Z}_2$ dimer liquid in an extended parameter regime (including the RK point) with  exponentially decaying dimer correlations and is characterized by a topological order in the form of four-fold ground state degeneracy for a system on a torus (as described in Fig.~\ref{fig:tor}).
The solid phases, on the other hand, spontaneously break various lattice symmetries of the Hamiltonian and thus have long range dimer-dimer correlations.
%

\begin{figure}
\begin{centering}
\includegraphics[width=8.0truecm]{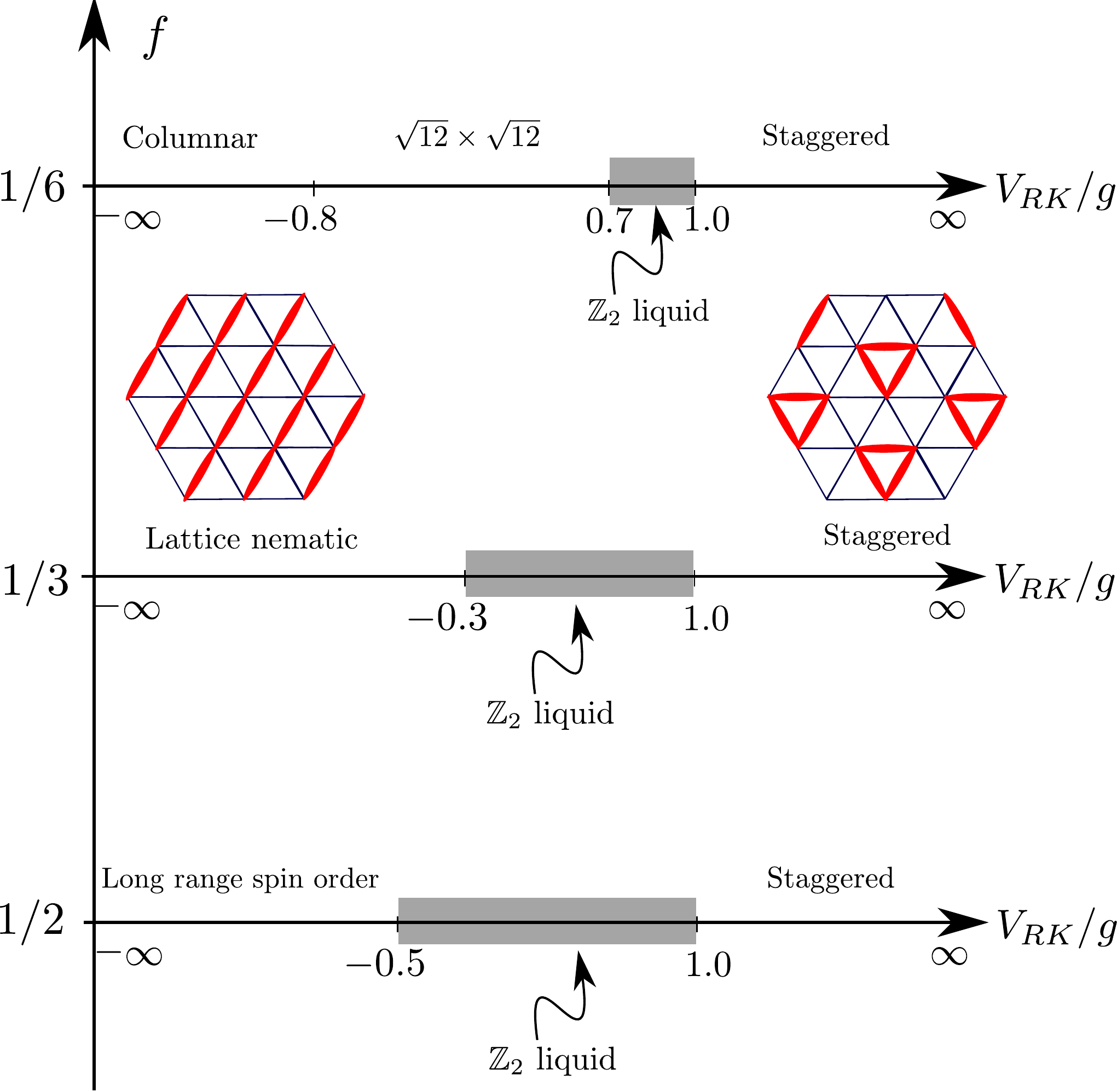}              
\end{centering}
\caption{(Color online) The phase diagram of the dimer models at different fractional fillings $f$. The important details about each of them are given in the text. The focus of the present paper is at $f=1/3$.
\label{fig:pd}}
\end{figure}

In context of the present work, as noted in the last section, we mention that the low energy effective Hamiltonian in the strong coupling limit generally assumes the RK form irrespective of the filling fraction of the bosons (or the equivalent dimer models: QDM, FPL or the 3-dimer model respectively for $f=1/6, 1/3$ and $1/2$) considered, albeit with important implications for the stability and nature of both the liquid and the solid phases which is summarized in Fig.~\ref{fig:pd}. With this general formulation we now specialize to the physics of the Mott lobe for $1/3$ filling.

\section{Numerical calculations at $1/3$ filling}
\label{sec_numerics}

We use a combination of tensor product states (TPS) formalism and exact diagonalization (ED) methods to obtain the phase diagram (Fig.~\ref{fig:pd} middle line) of the FPL model on the triangular lattice. 
Necessary details about implementing the numerical methods for different clusters are furnished in the following subsections facilitating a systematic analysis of our model. 
We take the value of $g$ to be 1 which is a convenient choice for further numerical calculations.


\subsection{Entanglement entropy at the RK point: the $\mathbb{Z}_2$ liquid}

\begin{figure}
\begin{centering}
\includegraphics[width=7truecm]{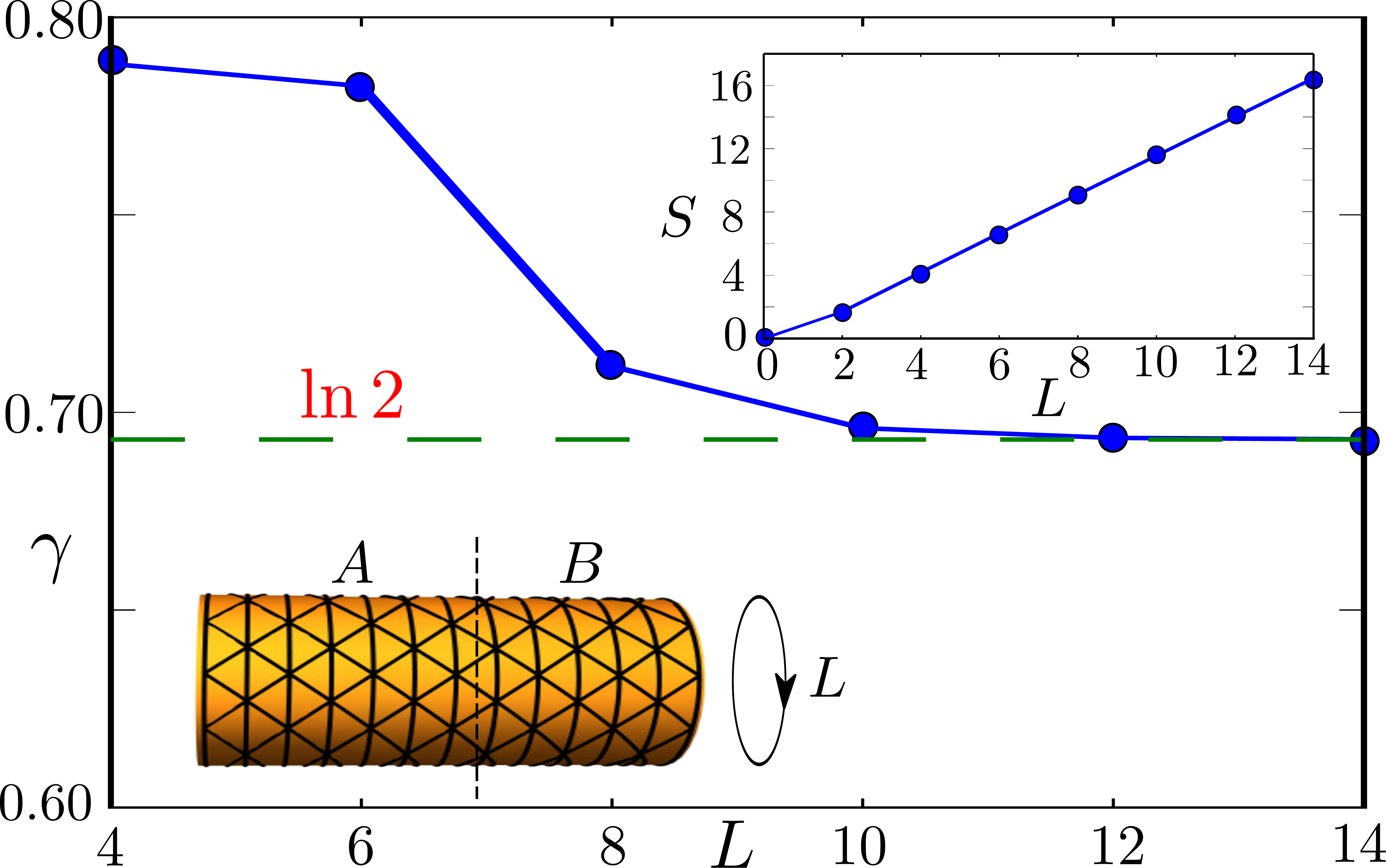}              
\end{centering}
\caption{(Color online) The plot shows the finite size dependence of the topological entanglement entropy $\gamma$ as a function of $L$, the perimeter of the subsystem $A$. The green dashed line corresponds to the saturated value of $\ln 2$ in the thermodynamic limit. The lattice is constructed on an infinite cylinder assuming periodic boundary condition in one direction. The black dashed line indicates the bipartition of the cylinder separating the subsystem $A$ from the rest $B$. The inset shows the linear growth of the entanglement entropy $S$ with $L$. 
\label{fig:renyi_entropy}}
\end{figure} 

One of the most interesting features of the Hamiltonian in Eq.~(\ref{tQDM}) is the existence of the exactly solvable RK point ($g=V_{\rm RK}$). We start by showing that at this point the ground state of our model (FPL) indeed has a topological order.

In order to characterize the topological order, an instructive quantity to look at is the entanglement entropy ($S$)  for a bipartition of the system into two parts $A$ and $B$ . 
The entanglement entropy of the reduced density matrix $\rho_A$ of subsystem $A$ is defined as $S = - \text{Tr} \left[\rho_A\log\rho_A\right]$ where $\rho_A$ is obtained from the full density matrix by tracing out all the degrees of freedom in the rest ($B$). 
For a gapped and topologically ordered gapped ground state, $S$ satisfies the ``area law"  which goes as,
\begin{equation}
 S(L) = \alpha L - \gamma + \mathcal{O}(L^{-1}) + \cdots ~,
 \label{entropy}
\end{equation}
where $\alpha$ in the leading term is a non-universal coefficient and $L$ is the perimeter of the subsystem A.
The sub-leading term $\gamma$, also known as the topological entanglement entropy,\cite{kitaev2006topological, levin2006detecting, amico2008entanglement, eisert2010colloquium} is, however, universal bearing the anyonic content of the state that reflects the topological order. 
This is directly related to the total quantum dimension (D) of the underlying topological field theory as $\gamma=\log{D}$.
Since $D = 2$ for a gapped $\mathbb{Z}_2$ liquid (described by the $\mathbb{Z}_2$ gauge theory), the quantity $\gamma$ in Eq.~(\ref{entropy}) saturates to $\log{2}$ in the thermodynamic limit i.e. $L\rightarrow \infty$.

The ground state at the RK point can be exactly represented by tensor networks using the framework of projected entangled-pair states (PEPS).\cite{verstraete2004renormalization, PhysRevB.83.245134}
We use this construction on a cylindrical triangular lattice (Fig.~\ref{fig:renyi_entropy}) to calculate $S$ as a function of $L$ and obtain $\gamma$ (Fig.~\ref{fig:renyi_entropy}) using Eq.~(\ref{entropy}).
The subsystem $A$ is constructed by bipartitioning the semi-periodic triangular lattice with the dashed line as shown in Fig.~\ref{fig:renyi_entropy}.
The circumference $L$ of the cylinder enters Eq.~(\ref{entropy}) as the perimeter of the subsystem and should be much larger than the maximum correlation length associated with the state.\cite{jiang2013accuracy}
The inset of Fig.~\ref{fig:renyi_entropy} shows the expected linear growth of $S$ with $L$ that is predicted by the leading term in Eq.~(\ref{entropy}).
The topological entanglement entropy ($\gamma$) is extracted from the intersection of the function $S$ on the $y$-axis when extrapolated backward and plotted against different values of $L$.
The tendency of $\gamma$ to saturate at the value of $\log{2}$ for large $L$ indicates to the fact that the RK point for the FPL model on the triangular lattice represents the ground state of a topologically ordered $\mathbb{Z}_2$ liquid akin to the other dimer models at $1/6$ and $1/2$ fillings.\cite{furukawa2007topological, isakov2011topological}
This is one of the main results of this work.

However, away from the RK point in the accessible parameter space, the ground state is no longer exactly known.
We adopt ED techniques to infer that the topological liquid is stable even away from the RK point over a finite window (see Fig.~\ref{fig:pd}) up till the system undergoes a quantum phase transition into the LN phase beyond a critical value of $V_{\text{RK}}$.
In the following subsections, we present the ED results containing the information about the low-lying spectrum of the model and measurements of various correlation functions that reflect the salient features of the phase diagram depicted in Fig.~\ref{fig:pd} middle line.

\subsection{Dimer-dimer correlation and the LN order}

\begin{figure}
\begin{centering}
\includegraphics[width=8.0truecm]{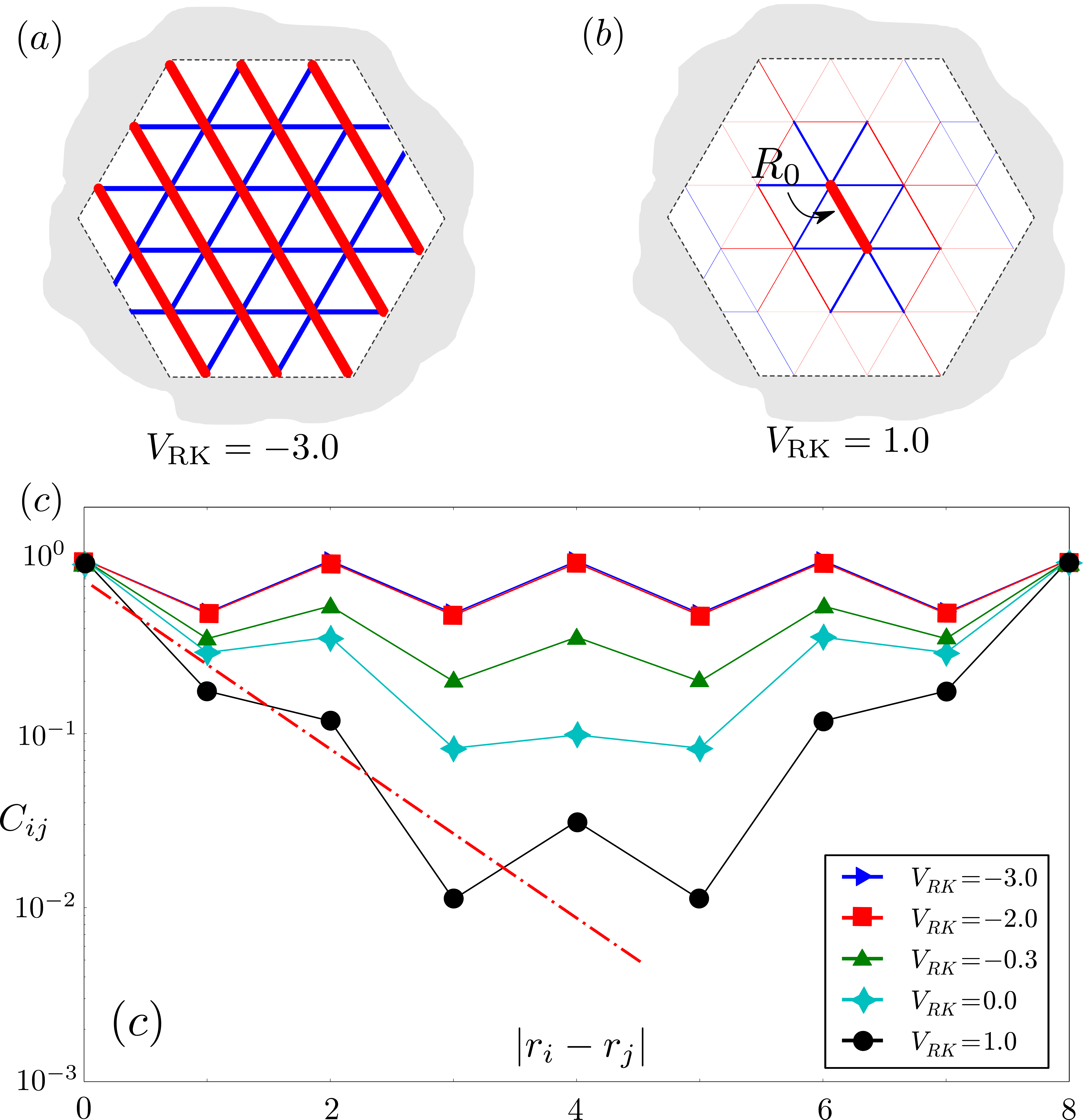}              
\end{centering}
\caption{(Color online) (a) The real space dimer-dimer correlation is plotted on a triangular lattice cluster at $V_{\text{RK}}=-3.0$ which indicates the parallel loop pattern deep into the LN phase. (b) Same is plotted for $V_{\text{RK}}=1.0$ showing the exponential decay of the dimer-dimer correlation function in the liquid phase at the RK point. The reference link, with respect to which correlations at all the other links are measured, is denoted by $R_0$. Red and blue links stand for positive and negative correlation respectively while the width measures the correlation strength. (c) Plotted is the same correlator as a function of $|r_{i}-r_{j}|$ at different RK potential. The red dashed line represents an approximate exponential fit $e^{-r/\xi}$ for $\xi\sim 1.0$. 
\label{fig:density_corr}}
\end{figure}

At the RK point, the loops are strongly fluctuating and the dimer-dimer correlations decay exponentially (as opposed to the algebraic decay on the square lattice) like the one in the QDM at $1/6$ filling. This indicates the lack of long range dimer order, as expected, in the $\mathbb{Z}_2$ liquid phase.
To check this, we calculate the dimer-dimer correlation function 
\begin{equation}
 C_{ij}(\vert{\bf r}_{i}-{\bf r}_{j}\vert)= \langle n({\bf r}_{i})n({\bf r}_{j}) \rangle - \langle n({\bf r}_{i}) \rangle \langle n({\bf r}_{j}) \rangle~,
\label{charge_corr}
\end{equation}
using ED on different clusters. Here $n({\bf r}_{i})$ is the dimer occupation number on $i^{\text{th}}$ link of the triangular lattice specified by the position vector ${\bf{r}}_i$. Each cluster has an extension of $L_x$ along the unit vector ${\bf{a}}_1=(1,0)$ and $L_y$ along the unit vector ${\bf{a}}_2=(\frac{1}{2},\frac{\surd3}{2})$ (see Fig.~\ref{fig:tor}). 

Although the considered clusters are small, the numerical results shown in Fig.~\ref{fig:density_corr}(c) suggest that the correlator (as a function of $r\equiv\vert{\bf r}_{i}-{\bf r}_{j}\vert$) is indeed exponential at the RK point with the correlation length $\xi\sim$ one lattice spacing. 
This is a known result for the case of the QDM on a triangular lattice (or equivalently the $1/6$ filling).\cite{PhysRevB.66.214513}
As mentioned earlier, at the RK point the model is exactly solvable and all the quantum correlations can be calculated classically (using some classical numerical techniques\cite{roychowdhury2014tensor}).   
The real space realization of the dimer correlation function at the RK point is shown in Fig.~\ref{fig:density_corr}(b) which points to an exponentially decaying nature of the correlation function as expected. 
The red and blue bonds correspond to the positive and negative correlation respectively with the width of the bond being proportional to the absolute value of the correlation function.  

The exponential decay disappears as $V_{\text{RK}}$ is taken to large negative values till we get into a phase with long range dimer-dimer correlations with $C(r)\simeq (-1)^r \frac{1}{4}$ (up to an offset) deep inside this phase signaling the onset of long range dimer order. 
To explore the nature of this long range ordered phase we plot the real space dimer-dimer correlation function in Fig.~\ref{fig:density_corr}(a). 
The plot suggests that the ordered phase is characterized by parallel alignments of loops on the triangular lattice which does not break any translation symmetry but the three-fold rotational symmetry (corresponding to the three-fold quasi-degeneracy in the ground sate of the spectrum).
Since the parallel loops do not have an orientation along the direction of their alignment, we refer to this as the LN phase.

\subsection{The phase transition between the $\mathbb{Z}_2$ liquid and the LN}

\begin{figure}
\begin{centering}
\includegraphics[width=8.0truecm]{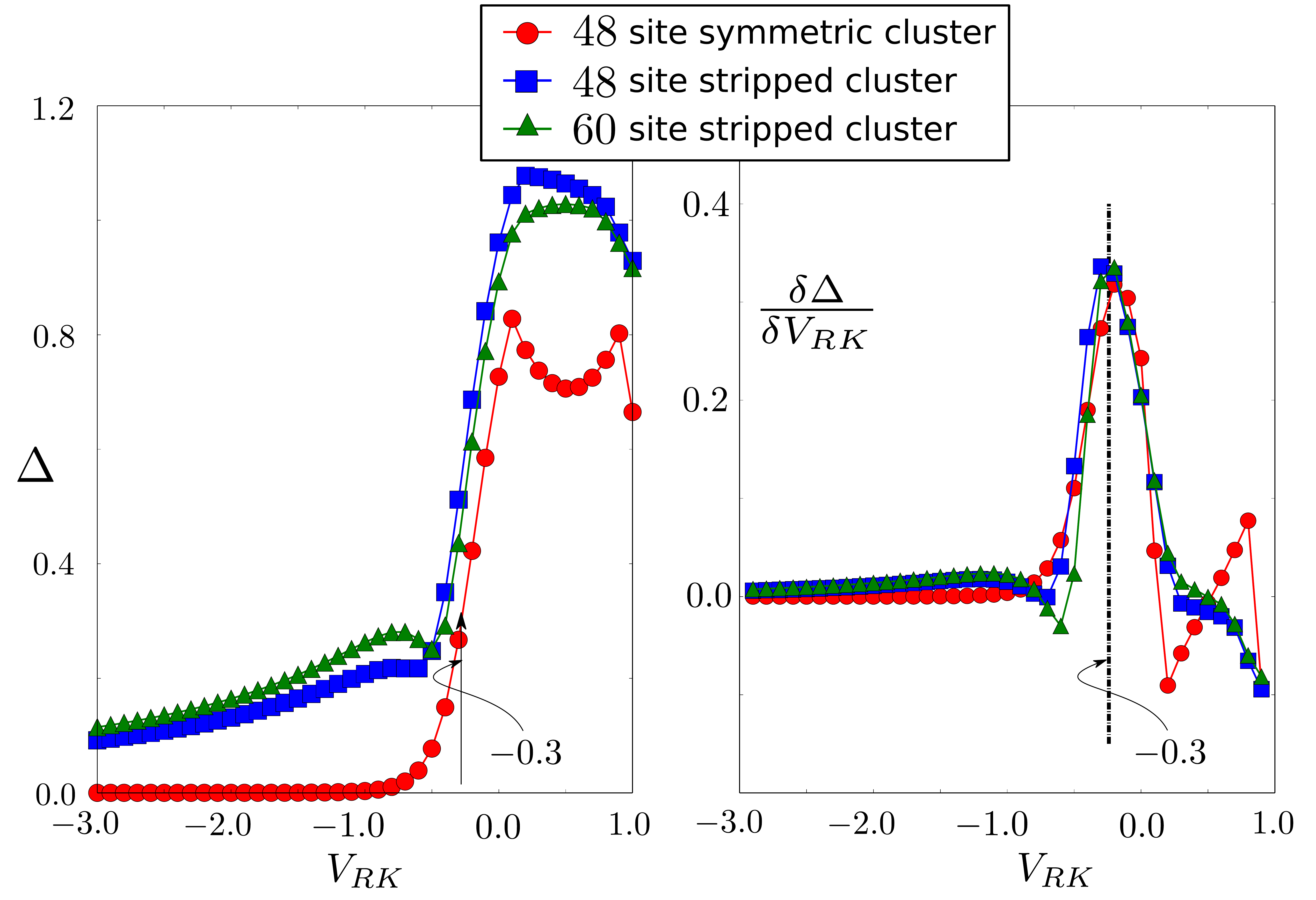}              
\end{centering}
\caption{(Color online) In the left panel, the gap to the first excited state tends to vanish at $V_{\text{RK}}\sim -0.3$ for different symmetric and stripped clusters. The ground state always lies in the zero-flux sector (0,0) for all of them. In the right panel, the derivative of the gap is plotted as a function of $V_{\text{RK}}$ to locate the transition point at $V_{\text{RK}}\sim-0.3$.  
\label{fig:gap_rk}}
\end{figure}

Having characterized the $\mathbb{Z}_2$ liquid phase (by non-trivial entanglement entropy and short range dimer-dimer correlations) and the LN phase (by long range dimer-dimer correlation), we now focus on the quantum phase transition between the two phases. 
First we study the excitation gap in the liquid with ED and observe that the gap closes only at a finite distance from the RK point (see Fig.~\ref{fig:gap_rk}). 
Thus the $\mathbb{Z}_2$ topological liquid persists over the whole region from $V_{\text{RK}}=1$ to $V_{\text{RK}}\sim-0.3$ which suggests that alone the kinetic term in Eq.~(\ref{tQDM}) can potentially stabilize the liquid phase even for $V_{\text{RK}}=0$.
The continuous vanishing of the gap near $V_{\text{RK}}=-0.3$ gradually destroys the liquid state driving the system into a charge-ordered LN phase as suggested from the behavior of the density-density correlation function shown in Fig.~\ref{fig:density_corr}(c). 

\begin{figure}
\begin{centering}
\includegraphics[width=8.0truecm]{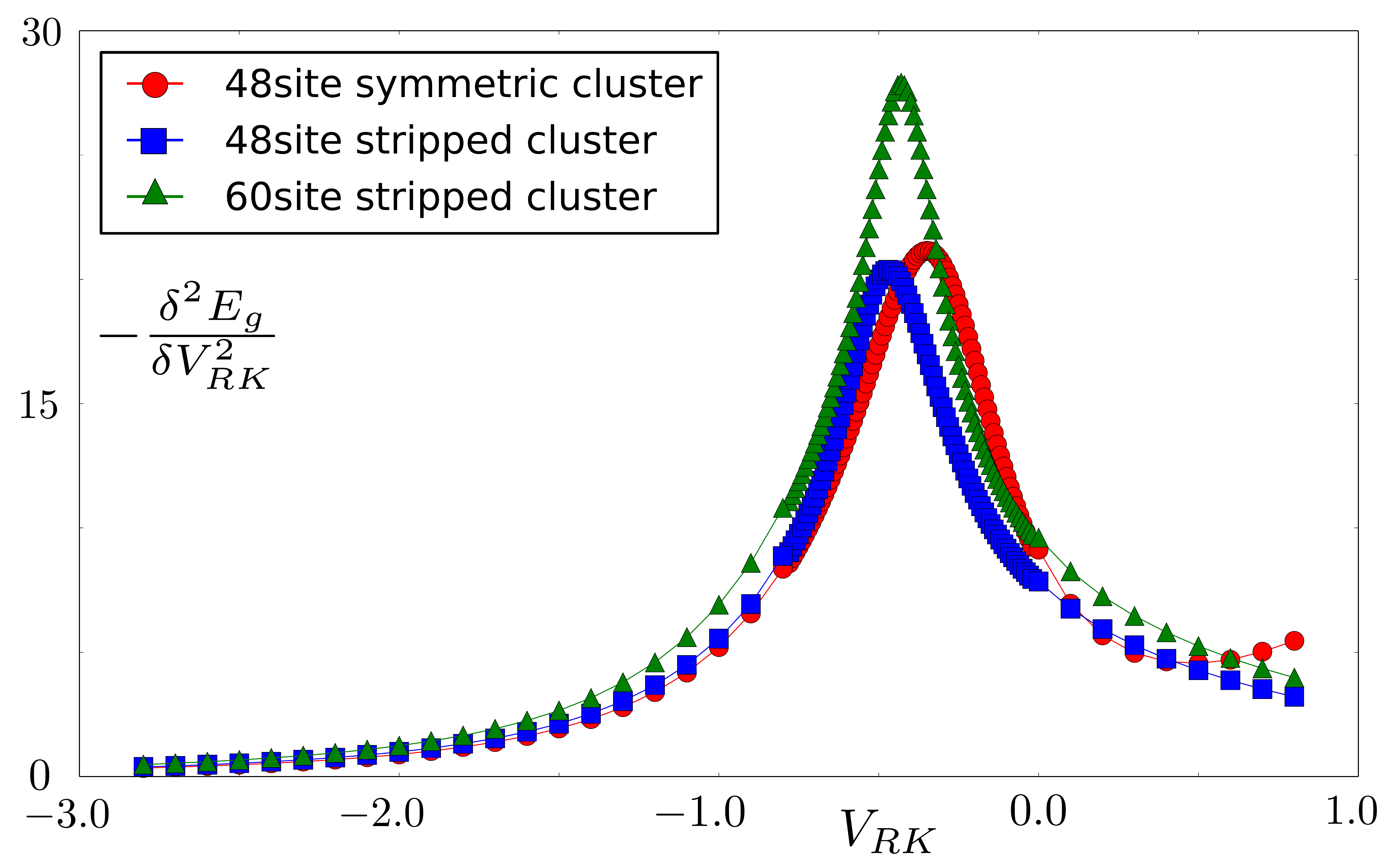}              
\end{centering}
\caption{(Color online) The ground state susceptibility $-\delta^{2}E_{g}/\delta V_{\text{RK}}^2$, in the (0,0) sector indicates a continuous phase transition between two phases for different clusters. The transition point is close to $V_{\text{RK}}\sim -0.3$ as evident from the right panel of Fig.~\ref{fig:gap_rk}. 
\label{fig:succep}}
\end{figure}

A generic way to locate the critical point within ED is to look at the response function of the system as a function of the parameters that define the Hamiltonian in Eq.~(\ref{tQDM}). 
In our case, an equivalent quantity can be framed using the second derivative of the ground state energy with respect to the RK potential: $-\delta^{2}E_{g}/\delta V_{\text{RK}}^2$. 
This is plotted in Fig.~\ref{fig:succep} where the single peak in the response is visible approximately at $V_{\text{RK}} \sim -0.3$. This shows that a single transition separates the topological liquid and the LN.
While it is impossible to rule out the possibilities of a first order transition from our finite cluster results, the smooth and gradual increase of the response along with the absence of shoulders suggests that the transition between the liquid and the LN may be continuous or weakly first order. 
%
%
The full energy spectrum of each of the clusters (not shown) suggests that at large negative $V_{\text{RK}}$, the ground state becomes nearly 3-fold degenerate (the quasi-degeneracy attributes to finite-size effect) where the three states are formed by superposition of the three loop patterns allowed by the $\mathcal{C}_3$ symmetry. 
We conclude on our numerical results from ED by noting that as the system gradually enters the ordered phase crossing the transition point, a set of three states in the bottom of the spectrum starts separating from the rest. Deep inside the LN phase, these three states become nearly degenerate (up to the finite size effects) with a finite excitation gap which is much higher in magnitude than the liquid gap and scales linearly with $V_{\text{RK}}$. This quasi-degeneracy is exact in the thermodynamic limit where the $\mathcal{C}_3$ symmetry is spontaneously broken.

Having established the two phases and the possibilities of a continuous phase transition between them, we now explore the critical theory for the predicted critical point (at $V_{\text{RK}}\sim-0.3$). We note that such a continuous transition would be very interesting in the sense that it describes the destruction of a topologically ordered phase.
%

\section{Continuous transition between the $\mathbb{Z}_2$ liquid and LN}
\label{sec_trans}

To construct a theory for the continuous phase transition between the $\mathbb{Z}_2$ liquid and the LN phases, we now introduce an  alternative spin representation of the hard-core boson model (or the equivalent dimer model).


\subsection{Spin representation and the gauge theory}

To obtain such a description, we first identify a spin $1/2$ degree of freedom on each site of the kagome lattice by virtue of the well known mappings: $b^\dagger_i=\sigma^+_i$, $b_i=\sigma^-_i$ and $n_i=(\sigma^z_i+1)/2$ where an up(down) spin represents presence(absence) of a boson at the lattice site.
Eq.~(\ref{potential}), then, becomes
\begin{equation}
 \mathcal{H}_{\text{V}} = V\sum_{\left\{ \mbox{\small\hexagon} \right\}} \big( \frac{1}{2} \sigma_{\mbox{\small\hexagon}}^z + h \big)^2 - \mu^2 N_{\mbox{\small\hexagon}}/16V~,
\label{potential_spin}
\end{equation}
where $N_{\mbox{\small\hexagon}}$ is the total number of kagome hexagons and $h=3-6f$.
The sum of all spin moments in a given hexagon is denoted by $S_{\mbox{\small\hexagon}}^z \equiv \frac{1}{2} \sigma_{\mbox{\small\hexagon}}^z$ where $ \sigma_{\mbox{\small\hexagon}}^z=\sum_{i\in{\hexagon}}\sigma^z_i$.
Clearly in the spin description, different fillings of bosons ($f$) correspond to different integer values of $h$ which essentially plays the role of an external magnetic field. 
Lowest energy configurations of the spin system specified by Eq.~(\ref{potential_spin}) satisfy the constraint that sum of the moments in every hexagon is exactly opposite to $h$ which depends on the filling factor $f$. 
Thus for $f=1/6, 1/3,1/2$; $h=2,1,0$ and hence potential term is satisfied if the total magnetization per hexagon at these fillings are $S^z_{\mbox{\small\hexagon}}=-2,-1,0$ respectively, corresponding to having one, two or three up spins (which means presence of one, two or three bosons as expected) per hexagon.

\subsection{Effective spin model}

In terms of the spins, the effective Hamiltonian (in Eq.~[\ref{h_eff}]) representing the dynamics within the degenerate ground state manifold of $\mathcal{H}_{\rm V}$ is given by, 
\begin{equation}
 \mathcal{H}_{\text{eff}} = \sum_{\alpha} \hat{\mathcal{P}}_{\alpha} \left[-g\prod_{\alpha}\sigma^x+ V_{\rm RK}\right].
 \label{eq_effective}
\end{equation}
Each term in Eq.~(\ref{eq_effective}) involves a product of four spins that form a bow-tie, as shown in Fig.~\ref{fig:bow}(a). The projector which selects out the flippable bow ties is expressed as\cite{balents2002fractionalization}
\begin{equation}
 \hat{\mathcal{P}}_{\alpha} = \sum_{\xi\pm1} \prod_{a\in \alpha} \left( \frac{1}{2} + \xi(-1)^{a} \sigma^z_a \right),
 \label{eq_projector}
\end{equation}
and we have also added the potential term in Eq.~(\ref{eq_effective}) to recover the RK Hamiltonian given in Eq.~(\ref{tQDM}). 

When the first term in Eq.~(\ref{eq_effective}) dominates,  the system prefers to align all the spins in the $\sigma^x$ direction and hence the boson number per site fluctuates. This is indeed the salient feature of a $\mathbb{Z}_2$ liquid phase (see below).  On the other hand when the second term dominates, the system prefers to choose a pattern to order in the $\sigma^z$ direction and we have a long range order in the boson density which turns out to be the LN phase that we discussed before. The actual magnitude of the coupling constants for which the transition between the two phases takes place depends on the details of the microscopic model.

\subsection{Ising gauge theory, visons and vison correlator}

\begin{figure}
\begin{centering}
\includegraphics[width=8.0truecm]{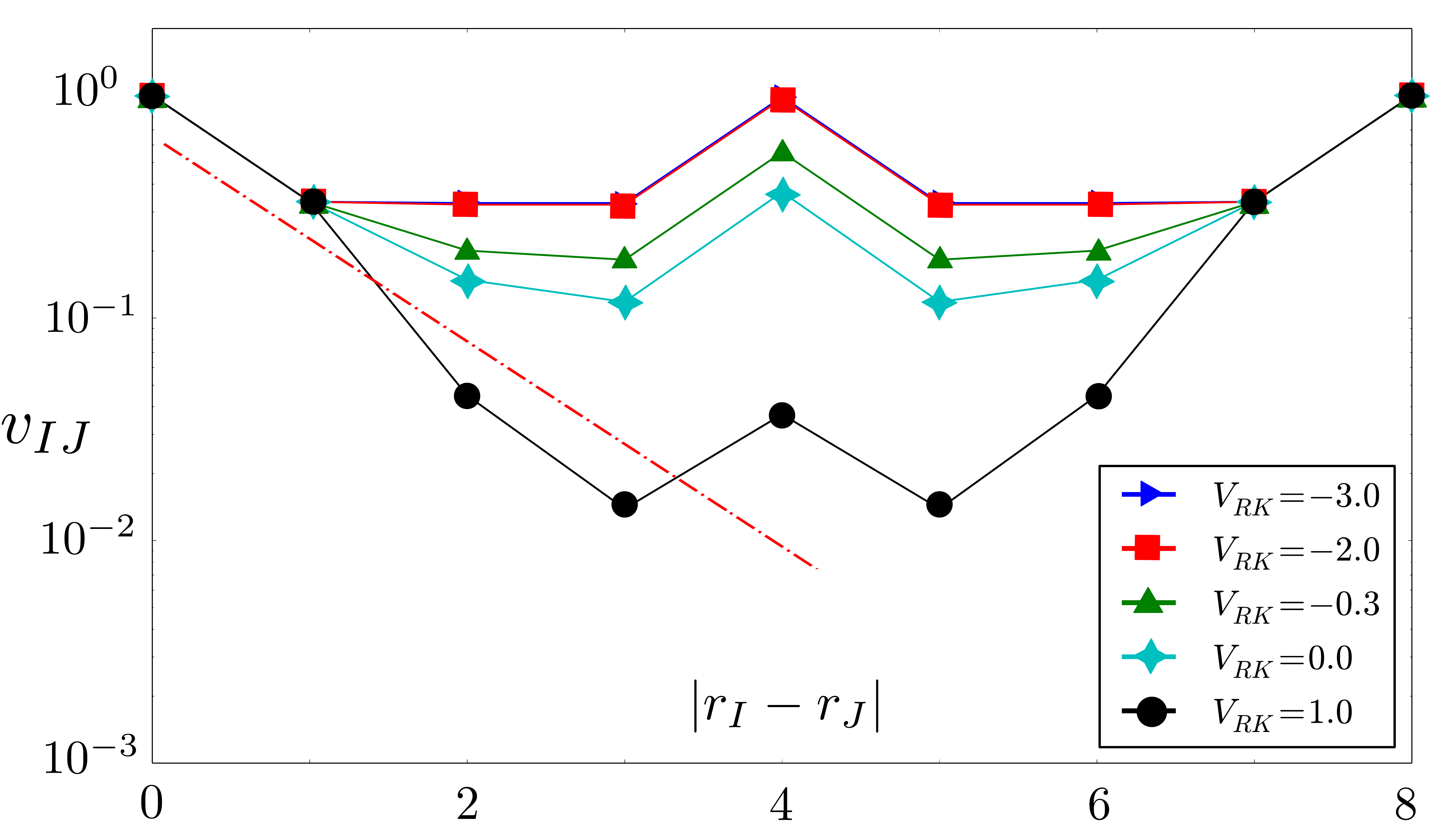}              
\end{centering}
\caption{(Color online) The two-vison correlator as a function of $\vert r_{I}-r_{J}\vert$ at different values of $V_{\text{RK}}$ for a symmetric cluster of $L=4$. The red dashed line represents an exponential decay: $e^{-\vert r_{I}-r_{J}\vert/\xi}$ for $\xi\sim 0.7$.
\label{fig:vison_corr}}
\end{figure}

Each spin sitting at the site of the kagome lattice is a part of two hexagonal plaquettes. We can define the Ising variables $\epsilon_{h}=\pm 1$ for each such hexagonal plaquette. Clearly the Hamiltonian in Eq.~(\ref{eq_effective}) is invariant under the $\mathbb{Z}_2$ gauge transformation\cite{balents2002fractionalization}
\begin{align}
\sigma^x_i\rightarrow \epsilon_h\sigma^x_i\epsilon_{h'}~,
\end{align}
where $h$ and $h'$ denotes the two hexagons of which the site $i$ is a part of. Such an Ising gauge structure, generated by the gauge transformations $\mathcal{G}_{h}=\exp[i\pi S_{\mbox{\small\hexagon}}^z]$, is an emergent property of the low energy subspace\cite{balents2002fractionalization} of the original microscopic bosonic Hamiltonian (in Eq.~[\ref{fullham}]) in the strong coupling limit ($V\gg t$). Indeed the above Hamiltonian represents an Ising gauge theory where the plaquette term $\mathcal{F}_\bigtriangleup=\left(\sigma^x_1\sigma^x_2\sigma^x_5\right)$ and $\mathcal{F}_\bigtriangledown=\left(\sigma^x_5\sigma^x_3\sigma^x_4\right)$, shown in Fig.~\ref{fig:bow}(a), measures the Ising magnetic flux through each triangle of the kagome lattice.

The operator $\sigma^z_i$ creates two such magnetic fluxes on the two triangles of the kagome lattice of which it is a part of. Such Ising flux excitations have been dubbed as {\it visons}.\cite{senthil2001frac} One can create a single vison excitation\cite{balents2002fractionalization} by applying a product of $\sigma^z$ operators along any path $C$ starting from any site of a given triangle of the kagome lattice to infinity:
\begin{align}
v^z_I=\prod_C^{I\rightarrow\infty} \sigma^z_i~,
\label{eq_vis}
\end{align}
where $i$ is a kagome lattice site encountered on the path $C$ which runs from the corresponding triangular kagome plaquette $I$, where the vison resides, to spatial $\infty$. Since the path operators {\it commute} with each other, it is straightforward to show that the two-vison wave function is symmetric under the exchange of the visons, or in other words, the visons are bosons themselves. With reference to the dimer covering, the above operator is nothing but the number of dimer variables encountered along the path. This immediately implies that the vison-vison correlator is given by\cite{ivanov2004vortexlike, strubi2011vison}
\begin{align}
\langle v^z_Iv^z_J\rangle=\langle \prod_C^{I\rightarrow J}\sigma^z_i \rangle \equiv v_{IJ} ~,
\end{align}
(where the path $C$ runs from $I$ to $J$) translates to
\begin{align}
\langle v^z_Iv^z_J\rangle=\langle (-1)^{N_{IJ}}\rangle~,
\label{vison}
\end{align}
where the operator $N_{IJ}$ counts the total number of dimers encountered on the path $C$ form $I$ to $J$ in any given dimer configuration. It is easy to verify that the explicite expressions of the vison operators in terms of the spins given in Eq.~(\ref{vison}) are independent of the contour $C$ up to an overall sign that can be fixed by measuring the correlator with respect to a fixed reference dimer configuration.

In the $\mathbb{Z}_2$ liquid phase, where $\mathcal{F}_\bigtriangleup=\mathcal{F}_\bigtriangledown=+1$, the vison excitations have a finite gap and the ground state does not contain free visons. On the other hand, the phase characterized by $\langle\sigma^z\rangle\neq0$, which we shall show is the LN phase, is a vison condensate.  
Thus we expect that the transition between the $\mathbb{Z}_2$ liquid and the LN is described by the closing of the vison gap leading to the vison condensation. 
Hence, the ground state expectation value of the vison correlator should also decay exponentially in the liquid phase with a length scale proportional to the inverse of the vison gap while it should have asymptotically reached a constant value in the LN phase. 

In Fig.~\ref{fig:vison_corr} we show the two-vison correlation function at different values of $V_{\text{RK}}$. 
The data at $V_{\text{RK}}=1.0$ fits well with the exponential curve $e^{-\vert r_{I}-r_{J}\vert/\xi}$ for $\xi\sim 0.7$. 
As $V_{\text{RK}}$ is decreased further, the system gradually enters the ordered phase and the two-vison correlator becomes asymptotically constant. This behavior is expected, as we shall show below.
Note that in Fig.~\ref{fig:vison_corr}, close to $V_{\text{RK}}=-0.3$ the value of $v_{IJ}$ changes by an order of magnitude much in the similar way as the density-density correlation does in Fig.~\ref{fig:density_corr}. 

\subsection{Lattice description for the visons}

Our numerical results indicate that the phase transition between the $\mathbb{Z}_2$ liquid and the LN phase, driven by the condensation of vison excitations, is possibly continuous. We shall now derive an effective critical theory for such a continuous transition. This would then compliment our numerical understanding of the phase diagram of the microscopic model.

\begin{figure}
\centering
\includegraphics[width=8.0truecm]{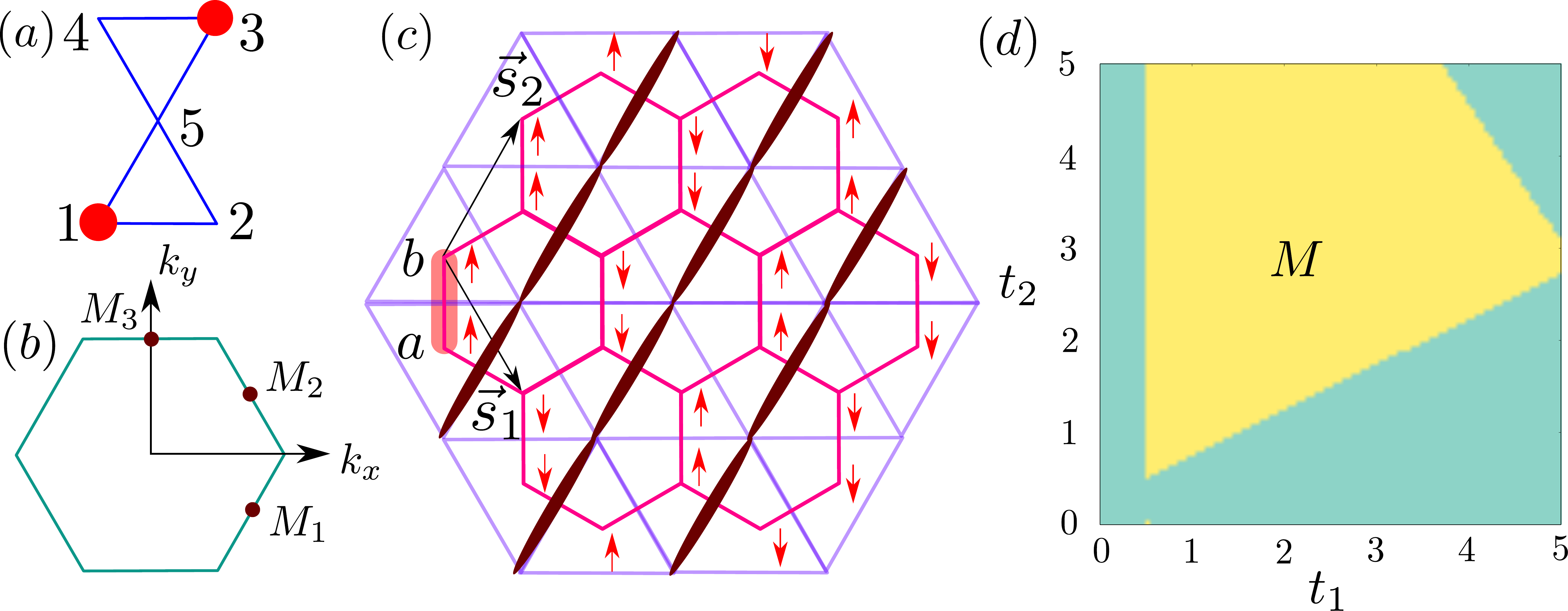}
\caption{(Color online) (a) A flippable bow tie. (b) The first Brillouin zone of the honeycomb lattice showing $M$ points. (c) Loop ordering at $M_1$ point. A loop segment is placed on the link of the triangular lattice whenever it crosses a honeycomb bond joining antiparallel spins which represent the vison degrees of freedom. (d) The yellow region qualitatively shows the allowed ranges of $t_1$ and $t_2$ for which the minima of $J(k)$ lie in the vicinity of the $M$ points within an energy window of $10^{-3}$. The vertical phase boundary indicates to the case when the minima are exactly on the $M$ points corresponding to the LN phase for $V_{\text{RK}}\ll0$.
\label{fig:bow}}
\end{figure}

In spirit of the universality of continuous phase transitions, to this end, we perform a series of mappings to isolate the vison degrees of freedom which we use to describe the critical theory for the transition. The effective Hamiltonian in Eq.~(\ref{eq_effective}) can be obtained from
\begin{align}
\tilde{\mathcal{H}}_{\rm eff}=&-g_{\rm eff}\sum_{\alpha} (\mathcal{F}_\bigtriangleup^{(\alpha)}+\mathcal{F}_\bigtriangledown^{(\alpha)})+V_{\rm eff}\sum_{\mbox{\small\hexagon}}\left(S_{\mbox{\small\hexagon}}^z+h\right)^2\nonumber\\
&+u_{\rm eff}\sum_{\alpha}\mathcal{P}_\alpha
\label{eq_eff}
\end{align}
in the limit of $g_{\rm eff}/V_{\rm eff},u_{\rm eff}/V_{\rm eff}\rightarrow 0$ $(g_{\rm eff},V_{\rm eff}>0)$ where the leading term is obtained in the second order perturbation theory with $g\sim g_{\rm eff}^2/V_{\rm eff}$ and $u_{\rm eff}\sim V_{\rm RK}$. We immediately note that the last two terms in Eq.~(\ref{eq_eff}) {\it commute} with each other as expected and hence, in the regime $|u_{\rm eff}|\gg g_{\rm eff}$ (and $V_{\rm eff}$ being the largest energy scale), the above model is rendered classical and the $u_{\rm eff}$ term chooses an appropriate order that is consistent with the filling ({\it i.e.} allowed by $V_{\rm eff}$). For $u_{\rm eff}<0$, this favors the order as depicted in Fig.~\ref{fig:bow}(c). 
On the other hand, when $|u_{\rm eff}|\ll g_{\rm eff}$, this order is expected to melt due to the quantum fluctuations in the spins. In particular, a state which allows flipping of spins in closed loops while maintaining the filling constraint becomes favorable. This is indeed consistent with the picture of the $\mathbb{Z}_2$ liquid of the microscopic model which is exact at the RK point. 

\begin{figure}
\begin{centering}
\includegraphics[width=8.7truecm]{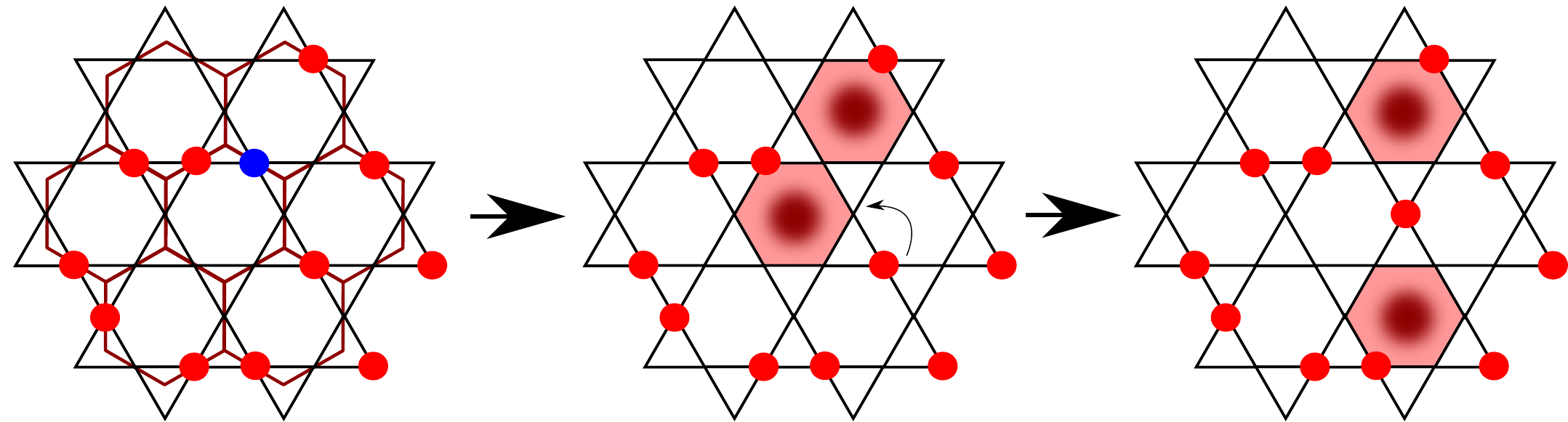}              
\end{centering}
\caption{(Color online) The medial honeycomb lattice is shown along with the original kagome lattice. Removal of the blue-colored boson creates two spinon excitations sitting at the centers of the adjacent shaded hexagons (the ``defect hexagons") which violate the constraint of having two bosons. By virtue of the single particle hopping term, these two defects can separate further apart, however their creation is energetically suppressed in the strong coupling limit.  
\label{fig:frac}}
\end{figure}

To arrive at the critical theory describing the vison condensation, we exploit the gauge structure, described in the last sub-section, of the effective theory to  isolate the vison degrees of freedom.
At this point, we introduce the honeycomb lattice which is the medial lattice of the kagome lattice as shown in the left most panel in Fig.~\ref{fig:frac}. On the sites of this medial lattice, we define the Ising variables $v^z=\pm 1$ and on its links we define the Ising gauge fields $\rho^z=\pm 1$.
The mapping from the $\sigma$ to the $v$ and $\rho$ variables is defined as follows,
\begin{equation}
 v_J^x = \prod_{\triangle} \sigma_a^x \sigma_b^x \sigma_c^x~;~~ \sigma_a^z = v_I^z \rho^z_{IJ} v_J^z~.
 \label{vison}
\end{equation}
Clearly, with the insight of Eq.~(\ref{eq_vis}), in the above mapping the $v^z_I$s are nothing but the visons (with conjugate momenta given by $v^x_I$). Since they carry Ising magnetic charge, they couple to the dual Ising gauge potential, $\rho^z_{IJ}$ and transform under a projective representation (that forms a projective symmetry group) of the symmetries of the underlying spin Hamiltonian. We note that the product of the dual gauge fields around the hexagonal plaquette is given by
\begin{equation}
 \prod_{\mbox{\small\hexagon}} \rho_{IJ}^z = \prod_{\mbox{\small\hexagon}} \sigma_a^z=\pm 1~,
 \label{eq_constraint}
\end{equation}
where the first product is over each honeycomb hexagon and the second one is on each kagome hexagon. 

Though not directly relevant to this work, we would like to point out that the presence of the dual magnetic flux, is nothing but the Ising electric charges ({\it spinons}):
\begin{equation}
 \prod_{\mbox{\small\hexagon}} \rho_{IJ}^z = \prod_{\mbox{\small\hexagon}} \sigma_a^z=- 1~,
\end{equation}
which sit at the center of the hexagonal plaquettes. In terms of the original bosons, these represent hexagons where the constraint of having two bosons is violated (see Fig.~\ref{fig:frac}). Evidently, such ``defect hexagons", are created in pairs and are energetically very costly in the strong coupling limit.  It is clear from the above equation that the visons see the spinons as a source of $\pi$-flux and hence the dual gauge potential naturally captures the mutual semionic statistics between the visons and the spinons.\cite{balents2002fractionalization} Now note that at $1/3$ filling, we have two bosons per hexagon. Hence, we must have

\begin{equation}
 \prod_{\mbox{\small\hexagon}} \rho_{IJ}^z = \prod_{\mbox{\small\hexagon}} \sigma_a^z=1~.
 \label{gauge_constrnt}
\end{equation}
This just means that there are no spinons at low energies because they are too costly. Hence, on circling the plaquettes of the medial honeycomb lattice, the ``flux'' seen by the $v^z$-spins is zero. This type of Ising gauge theory (IGT) are called even IGT as opposed to the odd IGT that arises in case of $1/6$ or $1/2$ filling when the $v^z$-spins sees a $\pi$-flux in each honeycomb plaquette.

With the mapping in Eq.~(\ref{vison}) the Hamiltonian in Eq.~(\ref{eq_eff}) becomes (we put $u_{\rm eff}=0$ for the moment)
\begin{align}
\mathcal{H}_{\rm dual}=-g_{\rm eff}\sum_I v^x_I+V_{\rm eff} \sum_{\mbox{\small\hexagon}}\left( \frac{1}{2}  \sum_{\langle I,J \rangle\in {\mbox{\small\hexagon}}}v^z_I\rho^z_{IJ}v^z_J+h\right)^2,
\label{eq_eff2}
\end{align}
expanding which we get,
\begin{align}
\mathcal{H}_{\rm dual}&=
-g_{\rm eff}\sum_I v^x_I + hV_{\rm eff}\sum_{\langle I,J \rangle}v^z_I\rho^z_{IJ}v^z_J  \nonumber \\
+\frac{V_{\rm eff}}{2} &\left[\sum_{\langle\langle II'\rangle\rangle}v^z_I\rho^z_{II'} v^z_{I'}+\sum_{I,J,I',J'\in{\mbox{\small\hexagon}}}v^z_I\rho^z_{IJ}v^z_Jv^z_{I'}\rho^z_{I'J'}v^z_{J'}\right]
\end{align}
up to a constant. Until now we have ignored the potential term in the RK Hamiltonian. This term, in the spin language, has the form
\begin{align}
V_{\rm RK}\sum_{\alpha}\mathcal{O}_{\alpha}
\end{align}
where $\alpha$ refers to all bow-ties and has the typical form, for the bow-tie in Fig.~\ref{fig:bow}(a), of
\begin{align}
\mathcal{O}_\alpha=&\frac{1}{8}\left[-\sigma^z_1\sigma^z_2-\sigma^z_3\sigma^z_4-\sigma^z_1\sigma^z_4-\sigma^z_2\sigma^z_3+\sigma^z_1\sigma^z_3+\sigma^z_2\sigma^z_4\right.\nonumber\\
&~~~~~~~~~~~~~~~~~~~~~~~~~~~~~~~~~~~~~~~~~~~~~~~~\left.-\sigma^z_1\sigma^z_2\sigma^z_3\sigma^z_4\right]~.
\end{align}
This, under the mapping to the $v_I$ variables, augments the second neighbor as well as the four-spin terms along with providing the interactions between four spins, two each on adjacent hexagons. As remarked earlier, the $v^z$ fields transform under a projective symmetry group (PSG) and in general a third neighbor Ising term for the $v^z_I$ fields would also be allowed by the PSG of an even-IGT. These term, would arise, for instance, on integrating out the four spin interactions which would also renormalize the nearest and the second neighbor interactions. In addition, a Maxwell term for the $\rho^z_{IJ}$ fields of the form $\prod_{\mbox{\small\hexagon}} \rho_{IJ}^z$ is also allowed and they renormalize the energy of the spinons, but due to the constraint in Eq.~(\ref{eq_constraint}), such terms are trivial and hence left out. Now, Because of the constraints in Eq.~(\ref{eq_constraint}) ({\it i.e.} no spinons), it is possible to choose a gauge where
\begin{align}
\rho^z_{IJ}=+1,~~~~\forall I,J~.
\label{even_gauge}
\end{align}
 Hence the general form of the model is given by
\begin{equation}
 \mathcal{H}_{\rm dual} = J_1\sum_{\text{n.n}} v_I^z v_J^z + J_2 \sum_{\text{n.n.n}} v_I^z v_J^z + J_3 \sum_{\text{n.n.n.n}} v_I^z v_J^z -\Gamma \sum_{J} v_I^x~.
 \label{vison_ham}
\end{equation}
Thus, the minimal gauge theory for the FPL model is  dual to the ferromagnetic transverse field Ising model on the honeycomb lattice with first, second and third neighbor Ising interactions.  

In the large $\Gamma$ limit we can neglect the $J$ terms and the $v$-spins are polarized in the $x$ direction with the finite vison gap $\sim2\Gamma$. The underlying $\sigma^z$ spins are thus fluctuating allowing for the boson number per site to fluctuate. This paramagnetic phase is nothing but the $\mathbb{Z}_2$ liquid in the dual description.

On increasing the Ising couplings ($J_{1,2,3}$), the visons gain dispersion and if the minima of the dispersion touches zero, they can condense leading to $\langle v^z_I\rangle\neq 0$. The nature of the ordering depends on the relative signs and magnitude of the Ising couplings and the phase diagram in the limit of large $J_{1,2,3}/\Gamma$ is shown in Fig.~\ref{fig:bow}(d). Setting $J_2/J_1=t_1$ and $J_3/J_1=t_2$ and $\Gamma/J_1=\tilde{\Gamma}$ with $J_1=1$, in Eq.~(\ref{vison_ham}), the soft vison modes\cite{blankschtein1984orderings, blankschtein1984fully} can be obtained from the Fourier transform of the Ising terms. This is given by

\begin{equation}
 \mathcal{H}_{\rm dual} = \sum_k J(k) \Psi_{k}^{\dagger} \Psi_k~,
\end{equation}
where $J(k)$ is the Fourier transform of the adjacency matrix of the Ising terms in Eq.~(\ref{vison_ham}): 
\begin{equation}
J(k) = \begin{pmatrix} 2t_1\delta & \gamma+t_2\eta \\ \gamma^*+t_2\eta^* & 2t_1\delta \end{pmatrix}
\label{J_k}
\end{equation}
and $\Psi_k=(v^x_{1k}~~v^x_{2k})^T$ and $\Psi_k^{\dagger}=(v^x_{1-k}~~v^x_{2-k})$. The momentum dependence of the parameters goes as
\begin{eqnarray}
 \delta &=& \cos{k_1} + \cos{k_2} + \cos{(k_1+k_2)} \nonumber\\
 \gamma &=& 1 + e^{i k_1} + e^{-i k_2} \nonumber\\
 \eta   &=& 2 \cos{(k_1+k_2)} + e^{i (k_1 - k_2)}~,
\end{eqnarray}
where $k_i={\vec{k}} \cdot {\vec{s}_i}$ with $\vec{s}_1=(\sqrt{3}/2,-3/2)$ and $\vec{s}_2 =(\sqrt{3}/2,3/2)$ being the basis vectors of the honeycomb lattice shown in Fig.~\ref{fig:bow}(c).
For an extended range of positive $t_1$ and $t_2$, as highlighted by the yellow region in Fig.~\ref{fig:bow}(d), the minima of the energy dispersion $J(k)$ occur at the three inequivalent $M$ points; $M_1$, $M_2$ and $M_3$ in the Brillouin zone [see Fig.~\ref{fig:bow}(b)] whose coordinates are ($\pi/\sqrt{3},-\pi/3$), ($\pi/\sqrt{3},\pi/3$) and ($0,2\pi/3$) respectively. This extended region of the phase diagram, where the ordering occurs at the $M$ points of the Brillouin zone of the medial honeycomb lattice, yields to a $v^z_I$ ordering pattern which is shown in Fig.~\ref{fig:bow}(c). Translating back (using Eq.~(\ref{vison}) and $n_i=(1+\sigma^z_i)/2$), this gives rise to the ordering pattern for $V_{\text{RK}}\ll0$. There are three such patterns (for the three $M$ points) which corresponds to the three LN phases related by the $\mathcal{C}_3$ symmetry as obtained from our previous numerical calculations.

While it is not possible to calculate the values of the Ising couplings in terms of the couplings of the underlying RK Hamiltonian in Eq.~(\ref{tQDM}), we observe that, the effective model allows phases that are observed in the microscopic model together with the possibilities of a direct continuous quantum phase transition from the $\mathbb{Z}_2$ liquid to the LN phase. Since the symmetries of the actual QDM and the effective gauge theories are identical, we expect that the transitions, which is attributed to the condensation of the visons, in both the models belong to the same universality class. We now attempt to find the structure of the critical theory which can predict the nature of this transition.

\subsection{Critical theory for the transition}

Approaching the transition point from the liquid side ( $\Gamma>J$ in Eq.~[\ref{vison_ham}]), we can write down the critical theory in terms of the vison modes that goes soft at the transition. 
These soft modes can be written as\cite{huh2011vison}
\begin{equation}
 \Psi(\vec{r}) = \sum_{j=1}^3 \psi_j(\vec{r}){\bf v}_j e^{i {\vec{M}_j}\cdot \vec{r}}~,
\end{equation}
where $\psi_j$ are the amplitudes of the three soft modes occurring at the three $M$ points of the Brillouin zone. To construct the Landau-Ginzburg action in terms of the soft modes we need to figure out the transformation of $\psi_j$ $(j=1,2,3)$ among themselves under various symmetries of the Hamiltonian. These are: (1) $\mathbb{T}_1$: lattice translation along $\vec{s}_1$, (2) $\mathbb{T}_2$: lattice translation along $\vec{s}_2$, (3) $\mathbb{I}$: bond inversion or parity,  (4) $\mathbb{C}_6$: rotation of $\pi/3$ about the center of a plaquette, and (5) global $\mathbb{Z}_2$ symmetry under which $v^x\rightarrow-v^x$. On a point of specific coordinates \{$x$,$y$\}, the actions of the above symmetries are the following: 
\begin{eqnarray}
 \mathbb{T}_1 &:& \{x,y;a,b \} \rightarrow \{x+1,y;a,b \} \nonumber \\
 \mathbb{T}_2 &:& \{x,y;a,b \} \rightarrow \{x,y+1;a,b \} \nonumber \\
 \mathbb{I}   &:& \{x,y;a,b \} \rightarrow \{-x,-y;b,a \} \nonumber \\
 \mathbb{C}_6 &:& \begin{cases}
                    \{x,y;a \} \rightarrow \{x-y+1,x;b\}, \\
                    \{x,y;b \} \rightarrow \{x-y,x;a \}.
                  \end{cases}
\end{eqnarray}
The transformation matrices of the three critical modes corresponding to different lattice symmetry operations and the global $\mathbb{Z}_2$ are as follows:
\begin{align}
 \mathcal{R}_{{\mathbb T}_1}          &= \begin{bmatrix} -1 & 0 & 0 \\ 0 & 1 & 0  \\ 0 & 0 & -1 \end{bmatrix} ~;~
 \mathcal{R}_{{\mathbb T}_2}           = \begin{bmatrix} 1 & 0 & 0  \\ 0 & -1 & 0 \\ 0 & 0 & -1 \end{bmatrix} ~;~ \nonumber \\
 \mathcal{R}_{\mathbb I}            &= \begin{bmatrix}  1 & 0 & 0 \\ 0 & 1 & 0  \\ 0 & 0 & -1 \end{bmatrix} ~;~~~~ 
 \mathcal{R}_{{\mathbb C}_6}           = \begin{bmatrix}  0 & 0 & -1 \\ 1 & 0 & 0  \\ 0 & 1 & 0  \end{bmatrix} ~;~  \nonumber \\
 \mathcal{R}_{\mathbb{Z}_2} &= \begin{bmatrix} -1 & 0 & 0  \\ 0 & -1 & 0 \\ 0 & 0 & -1 \end{bmatrix}.
\end{align}
These five matrices generate a 24 elements finite subgroup of $O(3)$ which is isomorphic to $C_2\otimes A_4$.\cite{gapgap}
Respecting all the above projective symmetry transformations, the most general Landau Ginzburg (LG) functional in $(2+1)$ dimensional Euclidian space-time assumes the following form,
\begin{align}
\mathcal{S}=\int d^2{\bf r}d\tau~\mathcal{L}~,
\end{align}
where the Lagrangian density (up to $6$th order) is given by
\begin{align}
 \mathcal{L} &= \nabla \vec\psi \cdot\nabla \vec\psi + \partial_{\tau} \vec\psi \cdot \partial_{\tau} \vec\psi  + r \vec\psi\cdot\vec\psi + {\tilde u}(\vec\psi\cdot\vec\psi)^2  \nonumber \\
 &+ {\tilde v}(\vec\psi\cdot\vec\psi)^3 + a (\psi_1^4 + \psi_2^4 +\vec\psi_3^4) + b (\psi_1\psi_2\psi_3)^2 
 \label{LG}
\end{align}
with $\psi = (\psi_1,~\psi_2,~\psi_3)^T$. If $a=b=0$, Eq.~(\ref{LG}) represents the usual soft spin $O(3)$ action (or the $N=3$ linear $\sigma$ model).
The $a$ term introduces cubic anisotropy in the system. 

Evidently, for $a<0$, the ordering occurs at one of the soft modes preferentially, {\it i.e.}, the functional is minimized when one of the components (among $\psi_1$, $\psi_2$, and $\psi_3$) takes a finite value while other two remain zero. 
Three such possibilities give rise to three symmetry oriented loop orderings. 
For example, the order at $M_1$, which corresponds to $\psi_1\ne0$ and $\psi_2 = \psi_3 = 0$, can be read from the structure of the full eigenvector ${\bf v}_1=(1,1)^T$ of $J(k)$ in Eq.~(\ref{J_k}) set to the momentum $M_1=(\pi/\sqrt{3},-\pi/3)$. 
The resultant $v^z$-spin configuration is shown in Fig.~\ref{fig:bow}(c) and the loop covering of the dimers can be obtained using Eq.~(\ref{vison}) which is equivalent to replacing the antiferromagnetic bonds of the honeycomb lattice by a loop segment on the underlying triangular lattice. This gives back the loop ordering which does not break any translation symmetry (because $M_1$ and $-M_1$ are identical and related by the momentum space lattice vectors) but spontaneously breaks the rotational symmetry. 
Similarly the ordering at other M points are related to the present one by the spontaneously broken $\mathcal{C}_3$ symmetry. This way the critical theory captures the ordering patterns obtained in the numerical calculations of the microscopic model. 

For $b=0$, leading order $\epsilon=4-d$ expansion suggests that the cubic anisotropy is irrelevant and the critical point is of $O(3)$ Wilson-Fisher type\cite{PhysRevB.8.4270, ketley1973modified, brezin1974discussion} (however, higher order expansion suggests that it may belong to the cubic critical point with critical exponents very close to the $O(3)$ class\cite{kleinert1995exact}). The $6$-th order anisotropy term (denoted by $b\neq 0$) is also irrelevant at such a point along with the $O(3)$ invariant $6$th order term, ${\tilde v}$ in Eq.~[\ref{LG}]. These considerations suggest that the phase transition between the $\mathbb{Z}_2$ liquid and the LN phase, as seen in the microscopic model, may belong to the $O(3)$ universality class and the anisotropy terms are dangerously irrelevant at this critical point.

It is important to note that the critical theory is not written in terms of the order parameter of the LN phase, as the conventional theory of phase transition would suggest.\cite{landau1974statistical, ma2000modern} Instead, it is naturally written in terms of vison fields and the order parameter is bilinear in terms of such vison fields. Hence we should expect large anomalous dimensions in the scaling dimension of the LN order parameter.\cite{sandvik2007evidence, isakov2012universal} Such large anomalous dimensions are characteristics to these types of unconventional phase transitions.

\section{summary and outlook}
\label{sec_summary}

In summary, we have studied the strong coupling limit of the extended Hubbard model for hard-core bosons with short range repulsive interactions.
Focusing on the particular fractional filling of $1/3$, we show that the low energy physics of the bosonic model is described by a quantum FPL model on the triangular lattice. Using a combination of different numerical techniques we analyze the ground state phase diagram of this quantum FPL model. 
Our numerical calculations conclusively establish the presence of a topologically ordered $\mathbb{Z}_2$ liquid phase over an extended parameter regime of the effective low energy Hamiltonian. 
On tuning appropriate parameters of the effective model, we find indication for a continuous phase transition from the topological phase to a phase that breaks the  $\mathcal{C}_3$ rotational symmetry of the triangular lattice. 
Using a mapping to an Ising gauge theory and PSG based arguments, we construct the effective field theory for a  generic continuous transition between the two phases, which we argue, belongs to the $O(3)$ universality class. 
We then show that such a transition is naturally related to the condensation of the (bosonic) magnetic charges of the Ising gauge theory, the so called visons. 
Hence, contrary to the conventional theory of continuous transition,\cite{ma2000modern} the present critical theory is written in terms of the soft vison modes and not the LN order parameter which is a bilinear in terms of the vison fields.

The present calculations show that the model of strongly interacting hard-core bosons can harbor rich and interesting phase diagrams including conventionally as well as topologically ordered states at different fractional fillings. 
It is interesting to note that considering the interacting particles to be fermions would induce a non-trivial statistics in the dimer problem. 
Interestingly, such kind of fermionic dimer models can arise in context of describing the metallic state of the hole-doped cuprates at low hole densities.\cite{punk2015quantum}
Whether a stable liquid phase can still be realized in such fermionic models requires more understanding regarding the underlying gauge theory and constitutes an interesting direction for further studies in future.

\section{Acknowledgments}

KRC would like to thank Sujit Das for critically reviewing the manuscript and SB thanks Roderich Moessner for helpful discussions.

\bibliography{reference_ordered}

\end{document}